\documentclass[preprint2,10.5pt]{modaastex}
\begin{document}

\title{\large{\rm{SPACE REDDENINGS FOR FIFTEEN GALACTIC CEPHEIDS}}}

\author{D.~G. Turner$^1$, R.~F. MacLellan$^2$, A.~A. Henden$^3$, L.~N. Berdnikov$^4$}
\affil{$^1$ Saint Mary's University, Halifax, Nova Scotia, Canada}
\affil{$^2$ Queen's University, Kingston, Ontario, Canada}
\affil{$^3$ AAVSO, Cambridge, Massachusetts, U.S.A.}
\affil{$^4$ Sternberg Astronomical Institute, Moscow, Russia}
\email{\rm{turner@ap.smu.ca}}

\begin{abstract}
Space reddenings are derived for 15 Galactic Cepheids from dereddening CCD {\it BV(RI)}$_C$ data for AF-type stars in the immediate vicinities of the variables, in conjunction with 2MASS reddenings for BAF-type stars in the same fields. Potential reddening solutions were analyzed using the variable-extinction method to identify stars sharing potentially similar distances and reddenings to the Cepheids, several of which have large color excesses. The intrinsic {\it BV(RI)}$_C$ color relation for AF dwarfs was modified slightly in the analysis in order to describe better the colors observed for unreddened stars in the samples. 
\end{abstract} 

\keywords{manuscript submitted for peer review.}

\section{{\rm \footnotesize INTRODUCTION}}
Because of the deleterious effects of interstellar reddening and extinction, it is difficult to establish an empirical picture of the Cepheid instability strip based entirely upon observations of Milky Way Cepheids. The use of extragalactic Cepheids is not necessarily a practical solution to the problem, since the available data for Milky Way Cepheids are generally more extensive and of greater precision, and the effects of internal reddening within other galaxies are not as well established as they are locally. Such considerations justify the continued use of the Galactic sample in studies aimed at establishing reliable intrinsic properties of classical Cepheid variables.

The determination of accurate reddenings for Cepheids has traditionally followed three different routes: (i) from field reddenings of specific objects (e.g., binary, cluster, association, or isolated Cepheids) based upon the analysis of photometric data for early-type stars sharing the same lines of sight, (ii) from observations mainly of bright Cepheids involving photometric parameters designed to be independent of interstellar reddening, and (iii) from using standard reddening laws and a calibrated intrinsic color relation (either observational or model generated) to deredden photometric observations for large samples of Cepheids. Methods (i) and (ii) are the most reliable means of establishing reddenings for individual Cepheids since method (iii) may entail use of period-color relations \citep[e.g.,][]{fe90a,fe90b}, which do not account for the intrinsic spread in effective temperature of the Cepheid instability strip and may generate erroneous results \citep[see][]{tu95}.

For method (ii), spectroscopic indices related to stellar effective temperature and designed to be independent, or relatively independent, of interstellar and atmospheric extinction include the $\Gamma$-index \citep{kr60,sj89}, which measures the depression of the {\it G}-band of CH ($\lambda$4305) relative to the local continuum, the $\beta$-index, which samples the strength of the H$\beta$ Balmer line of hydrogen relative to the surrounding continuum, and the {\it KHG}-index of Brigham Young University \citep{mp69,mc70,fm72}, which measures the strengths of Ca {\sc II} K ($\lambda$3933), Balmer H$\delta$ ($\lambda$4101), and the {\it G}-band through narrow band interference filters in a manner independent of extinction. A more recent technique \citep{sl90} involves the use of line depth ratios between C {\sc I} and Si {\sc II} lying on the Brackett continuum near 10728 \AA. A similar technique was used by \citet{kr98} using spectral lines falling in the red region of Cepheid spectra. \citet{ko08} have taken the technique to its ultimate level of sophistication by using high resolution optical spectra of Cepheids throughout their cycles in conjunction with stellar atmosphere models to track their changes in effective temperature, and thus intrinsic broad band color. 
 
The use of close neighbours to deduce reddenings for Cepheids appears to date from its application to the blue companion of $\delta$ Cep by \citet{eg51}; it has also been used recently for Cepheids studied with the Hubble Space Telescope \citep{be07}. But the existing sample includes only $\sim40$ Cepheids of known space reddening \citep{lc07}. This study presents new space reddenings for 15 Galactic Cepheids derived from CCD {\it BV(RI)}$_C$ photometry for stars in the immediate fields of the variables. The goal was to test the feasibility of deriving Cepheid reddenings based upon only {\it BV(RI)}$_C$ observations for stars in the fields of the Cepheids, as well as to augment the limited sample of Cepheids with space reddenings. As demonstrated here, the methodology appears to be provide useful results that extend the sample of Cepheids with independently derived reddenings.

\section{{\rm \footnotesize OBSERVATIONAL DATA}}
The input data for the present study consist of observations of 15 Cepheid fields obtained with the 1.0-m Ritchey-Chr\'{e}tien telescope of the U.S. Naval Observatory, Flagstaff Station. A Tektronix/SITe, 1024 $\times$ 1024 pixel, thinned CCD was used with Johnson system {\it BV} and Kron-Cousins system {\it (RI)}$_C$ filters to image the fields \citep[e.g.,][]{hm00}. Several deep images were obtained for each field, from which average {\it BV(RI)}$_C$ magnitudes and colors, with typical uncertainties smaller than $\pm0^{\rm m}.01$, were extracted for all detectable stars using DAOPHOT \citep{st87}. Since no {\it U}-band observations or spectral types are available for the stars in the survey fields, it was necessary to deduce individual reddenings from photometric analyses of two-color diagrams constructed from the {\it BV(RI)}$_C$ data. The data are available electronically from A.A.H. via the AAVSO.\footnote{\url{ftp://www.aavso.org/public/calib/}}

An often-ignored feature of interstellar reddening is that no single relationship accurately describes it over all regions of the Galaxy \citep{wa61,wa62,ma90,tu76a,tu76b,tu89,tu94}. Regional variations in the reddening law depend directly upon direction viewed through the Milky Way, vary slowly with Galactic longitude, and are tied to differences in the distribution of particle sizes for dust lying along different lines of sight. The resulting Galactic longitude and latitude dependence of the extinction law affects interstellar reddening studies that implicitly use a relationship of fixed parameterization to describe the extinction in a particular color system, and should be most important for stars of large reddening. Several of the Cepheids in the present sample are heavily reddened, so the extinction laws for the fields of each object were established prior to the analysis by taking advantage of existing studies for regions reasonably close to the program objects \citep[e.g.,][as well as other published studies by the lead author]{jo68,tu76b,tu89}.

\begin{figure}[!t]
\begin{center}
\includegraphics[width=7.0cm]{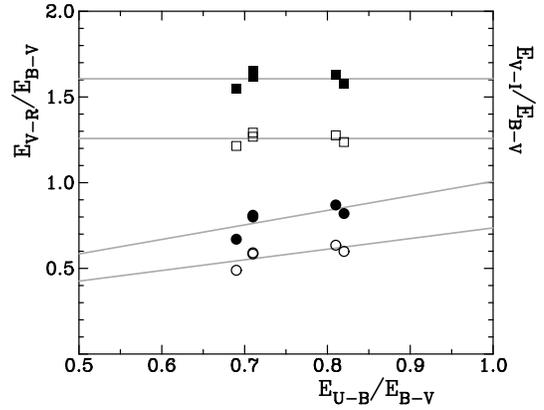}
\end{center}
\caption{\small{Derived Johnson system (filled symbols) and Kron-Cousins system (open symbols) reddening ratios $E_{V-R}/E_{B-V}$ (circles, lower) and $E_{V-I}/E_{B-V}$ (squares, upper) for fields studied by Johnson (1968). Gray lines denote the implied dependences on reddening slope $E_{U-B}/E_{B-V}$.}}
\label{fig1}
\end{figure}

Specifically, a reddening slope $E_{U-B}/E_{B-V}$ for the field of each program Cepheid was established from studies by \citet{tu76b,tu89} for adjacent regions, and was linked to specific reddening slopes $E_{V-I}/E_{B-V}$ and $E_{R-I}/E_{B-V}$ on the Johnson system using regional reddening curves from \citet{jo68}. Those were then converted to reddening slopes $E_{V-I}/E_{B-V}$ and $E_{R-I}/E_{B-V}$ on the Kron-Cousins system using the results of \citet{fe83} and \citet{ca93}. To illustrate the methodology, the results for specific fields studied by \citet{jo68} and exhibiting a range of color excess ratios $E_{U-B}/E_{B-V}$ are shown in Fig.~\ref{fig1}, where it can be noted that the {\it I--}band $E_{V-I}/E_{B-V}$ reddening ratios appear to display no significant dependence on visible reddening slope $E_{U-B}/E_{B-V}$. The Cepheids studied here all lie in fields where the reddening slope does not vary significantly from the locally observed Galactic mean, so specifying the exact reddening ratio for each field proved to be only a minor concern.

\section{{\rm \footnotesize METHOD OF ANALYSIS}}
\begin{figure}[!t]
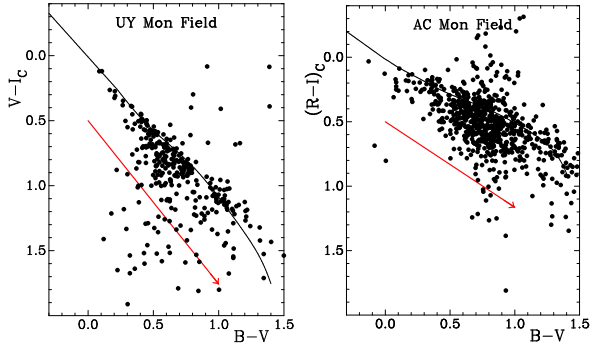

\begin{center}
\includegraphics[width=1.5in]{f2a.eps}
\includegraphics[width=1.5in]{f2b.eps}
\end{center}
\caption{\small{Two-color diagrams, {\it V--I}$_C$ versus {\it B--V} for the field of UY Mon (left), and {\it (R--I)}$_C$ versus {\it B--V} for the field of AC Mon (right), showing the intrinsic color relations adopted (solid curves). Reddening lines specific for the fields are shown separately in red for $E_{B-V} = 1.0$, with arrows denoting the direction of increasing reddening.}}
\label{fig2}
\end{figure}

This study presents new space reddenings for 15 Galactic Cepheids derived from two-color diagrams, {\it V--I}$_C$ versus {\it B--V} and {\it (R--I)}$_C$ versus {\it B--V}, constructed from CCD {\it BV(RI)}$_C$ photometry for stars in the fields of the variables. Reddening lines appropriate for each field were adopted as indicated above, specifically the adopted reddening slopes $E_{U-B}/E_{B-V}$ were 0.73 for VY Sgr, AY Sgr, and V1882 Sgr, 0.75 for HZ Per and OT Per, 0.76 for FO Cas, IO Cas, EW Aur, YZ CMa, CN CMa, BD Pup, BE Pup, and LR Pup, and 0.77 for UY Mon and AC Mon. The reddening slopes $E_{V-I}/E_{B-V}$ and $E_{R-I}/E_{B-V}$ were next obtained through interpolation and computation with the data of Fig.~\ref{fig1}, resulting in values on the Kron-Cousins system of $E_{V-I}/E_{B-V}  = 1.257$ for all fields and values of $E_{R-I}/E_{B-V}$ of 0.689, 0.677, 0.670, and 0.664, respectively, for the Cepheid fields listed above. Examples are shown in Fig.~\ref{fig2}, where the derived reddening slopes for UY Mon and AC Mon are $E_{V-I}/E_{B-V} = 1.257$ and $E_{R-I}/E_{B-V} = 0.664$ according to the inferred reddening ratio $E_{U-B}/E_{B-V} = 0.73$. Note that the adopted field reddening lines are unchanging for $E_{V-I}/E_{B-V}$ and exhibit only small variations for $E_{R-I}/E_{B-V}$.

Initial tests with two-color diagrams using intrinsic {\it BV(RI)}$_C$ colors for dwarfs from \citet{ca93} revealed small anomalies in the inferred $E_{B-V}$ reddenings derived from the two diagrams, as well as an overabundance of unreddened and negatively-reddened stars relative to the intrinsic relation, not explainable as luminosity effects \citep{ca93}. Such non-physical results suggested the need for slight corrections to the intrinsic colors used with the present data sets, which are normalized to the Kron-Cousins system. Small adjustments to the intrinsic relations for late B-type and A-type dwarfs to make them bluer were therefore tried using alternative intrinsic colors from \citet{jo66} adjusted to the Cape system \citep{fe83}. That produced greater consistency in the derived reddenings and better agreement for unreddened stars, and was adopted throughout the remainder of the study. The intrinsic colors corresponding to such changes are presented in Table~\ref{tab1} for reference purposes, where they are compared with the \citet[][SAAO]{ca93} colors for dwarfs, which were adopted for all other stars. Given that the analysis was restricted to stars that dereddened uniquely to the AF-dwarf relation, the modification affects the resulting space reddenings, but only to a minor extent. Likely B-type stars in the field of each Cepheid were only used when analyzing 2MASS colors for the stars.

\setcounter{table}{0}
\begin{table}[!t]
\caption[]{Intrinsic {\it BV(RI)}$_C$ Colors for Dwarfs.}
\label{tab1}
\centering
\tiny
\begin{tabular*}{0.47\textwidth}{@{\extracolsep{-3.7mm}}ccccccccccc}
\hline \noalign{\medskip}
{\it (B--V)}$_0$ &({\it V--I})$_0$ &({\it V--I})$_0$ &({\it R--I})$_0$ &({\it R--I})$_0$ & &{\it (B--V)}$_0$ &({\it V--I})$_0$ &({\it V--I})$_0$ &({\it R--I})$_0$ &({\it R--I})$_0$ \\
& &SAAO & &SAAO & & & &SAAO & &SAAO \\
\noalign{\smallskip} \hline \noalign{\medskip}
\noalign{\smallskip}
$-0.24$ &$-0.259$ &$-0.237$ &$-0.157$ &$-0.158$ & &$+0.04$ &$+0.055$ &$+0.037$ &$+0.040$ &$+0.021$ \\
$-0.23$ &$-0.248$ &$-0.221$ &$-0.150$ &$-0.146$ & &$+0.05$ &$+0.066$ &$+0.048$ &$+0.046$ &$+0.026$ \\
$-0.22$ &$-0.236$ &$-0.206$ &$-0.143$ &$-0.134$ & &$+0.06$ &$+0.077$ &$+0.059$ &$+0.053$ &$+0.032$ \\
$-0.21$ &$-0.225$ &$-0.193$ &$-0.136$ &$-0.123$ & &$+0.07$ &$+0.088$ &$+0.070$ &$+0.059$ &$+0.038$ \\
$-0.20$ &$-0.214$ &$-0.181$ &$-0.129$ &$-0.114$ & &$+0.08$ &$+0.100$ &$+0.081$ &$+0.066$ &$+0.043$ \\
$-0.19$ &$-0.203$ &$-0.170$ &$-0.121$ &$-0.105$ & &$+0.09$ &$+0.111$ &$+0.093$ &$+0.072$ &$+0.049$ \\
$-0.18$ &$-0.192$ &$-0.159$ &$-0.114$ &$-0.097$ & &$+0.10$ &$+0.122$ &$+0.105$ &$+0.078$ &$+0.055$ \\
$-0.17$ &$-0.180$ &$-0.149$ &$-0.107$ &$-0.089$ & &$+0.11$ &$+0.133$ &$+0.117$ &$+0.084$ &$+0.061$ \\
$-0.16$ &$-0.169$ &$-0.140$ &$-0.100$ &$-0.082$ & &$+0.12$ &$+0.144$ &$+0.129$ &$+0.090$ &$+0.068$ \\
$-0.15$ &$-0.158$ &$-0.131$ &$-0.092$ &$-0.076$ & &$+0.13$ &$+0.156$ &$+0.141$ &$+0.095$ &$+0.074$ \\
$-0.14$ &$-0.147$ &$-0.122$ &$-0.085$ &$-0.070$ & &$+0.14$ &$+0.167$ &$+0.153$ &$+0.101$ &$+0.080$ \\
$-0.13$ &$-0.136$ &$-0.114$ &$-0.078$ &$-0.064$ & &$+0.15$ &$+0.178$ &$+0.166$ &$+0.106$ &$+0.086$ \\
$-0.12$ &$-0.124$ &$-0.106$ &$-0.071$ &$-0.059$ & &$+0.16$ &$+0.189$ &$+0.179$ &$+0.112$ &$+0.093$ \\
$-0.11$ &$-0.113$ &$-0.098$ &$-0.064$ &$-0.053$ & &$+0.17$ &$+0.200$ &$+0.191$ &$+0.117$ &$+0.099$ \\
$-0.10$ &$-0.102$ &$-0.090$ &$-0.056$ &$-0.048$ & &$+0.18$ &$+0.212$ &$+0.204$ &$+0.122$ &$+0.106$ \\
$-0.09$ &$-0.091$ &$-0.082$ &$-0.049$ &$-0.043$ & &$+0.19$ &$+0.223$ &$+0.217$ &$+0.127$ &$+0.112$ \\
$-0.08$ &$-0.080$ &$-0.074$ &$-0.042$ &$-0.039$ & &$+0.20$ &$+0.234$ &$+0.230$ &$+0.133$ &$+0.119$ \\
$-0.07$ &$-0.068$ &$-0.066$ &$-0.035$ &$-0.034$ & &$+0.21$ &\nodata &$+0.243$ &$+0.138$ &$+0.125$ \\
$-0.06$ &$-0.057$ &$-0.057$ &$-0.028$ &$-0.029$ & &$+0.22$ &\nodata &$+0.256$ &$+0.143$ &$+0.132$ \\
$-0.05$ &$-0.046$ &$-0.049$ &$-0.020$ &$-0.024$ & &$+0.23$ &\nodata &$+0.269$ &$+0.148$ &$+0.138$ \\
$-0.04$ &$-0.035$ &$-0.040$ &$-0.013$ &$-0.020$ & &$+0.24$ &\nodata &$+0.281$ &$+0.154$ &$+0.144$ \\
$-0.03$ &$-0.024$ &$-0.031$ &$-0.006$ &$-0.015$ & &$+0.25$ &\nodata &$+0.294$ &$+0.159$ &$+0.151$ \\
$-0.02$ &$-0.012$ &$-0.022$ &$+0.001$ &$-0.010$ & &$+0.26$ &\nodata &$+0.307$ &$+0.164$ &$+0.157$ \\
$-0.01$ &$-0.001$ &$-0.013$ &$+0.008$ &$-0.005$ & &$+0.27$ &\nodata &$+0.320$ &$+0.170$ &$+0.164$ \\
$+0.00$ &$+0.010$ &$-0.004$ &$+0.015$ &$+0.000$ & &$+0.28$ &\nodata &$+0.332$ &$+0.175$ &$+0.170$ \\
$+0.01$ &$+0.021$ &$+0.006$ &$+0.021$ &$+0.005$ & &$+0.29$ &\nodata &$+0.345$ &$+0.180$ &$+0.176$ \\
$+0.02$ &$+0.032$ &$+0.016$ &$+0.027$ &$+0.010$ & &$+0.30$ &\nodata &$+0.357$ &$+0.185$ &$+0.182$ \\
$+0.03$ &$+0.043$ &$+0.026$ &$+0.033$ &$+0.016$ \\
\noalign{\smallskip} \hline
\end{tabular*}
\end{table}

Reddening lines run nearly parallel to the intrinsic relations for B-dwarfs and KM-dwarfs, particularly in {\it V--I}$_C$ versus {\it B--V} diagrams, which is why the analysis was restricted to stars indicated to be likely AF-dwarfs, which are reasonably plentiful in each field. The slope of the intrinsic relation for AF-dwarfs relative to the effects of interstellar reddening in the {\it BV(RI)}$_C$ system results in sufficient separation, particularly in {\it (R--I)}$_C$ relative to {\it B--V}, to produce unique photometric dereddening solutions for each star, and the use of two separate color-color diagrams provides independent estimates for the intrinsic color of each star, although greater precision is obtained with solutions from the {\it (R--I)}$_C$ versus {\it B--V} diagram alone, because the angle between the intrinsic relation and typical reddening lines is larger. Infrared {\it JHK}$_{\rm s}$ observations exist for most stars \citep{cu03} from the Two Micron All Sky Survey \citep[2MASS,][]{sk06}, and confirm the adopted reddenings from {\it BV(RI)}$_C$ data, although the scatter in 2MASS {\it JHK}$_{\rm s}$  colors tends to be rather significant, larger than in {\it UBV} color-color diagrams \citep{tu76a}.

Typical companions and progenitors of Cepheids are B-type stars \citep{tu84}, which are rare enough that their occurrence in the field of a Cepheid raises the possibility of a physical association. AF-type stars, on the other hand, are a more common constituent of Galactic star fields \citep{mc65}, so the possibility of their physical association with a nearby Cepheid is reduced, but not necessarily to zero. In many of our program fields some of the stars identified as likely AF-type may be associated with the Cepheid of interest, but that was not explored here since there are no catalogued star clusters in the fields, although the regions around UY Mon, BE Pup, YZ CMa, and VY Sgr appear to contain faint anonymous clusters, and BD Pup and FO Cas are located in bright groups of surrounding stars.

\begin{figure*}[!t]
\begin{center}
\includegraphics[width=10.0cm]{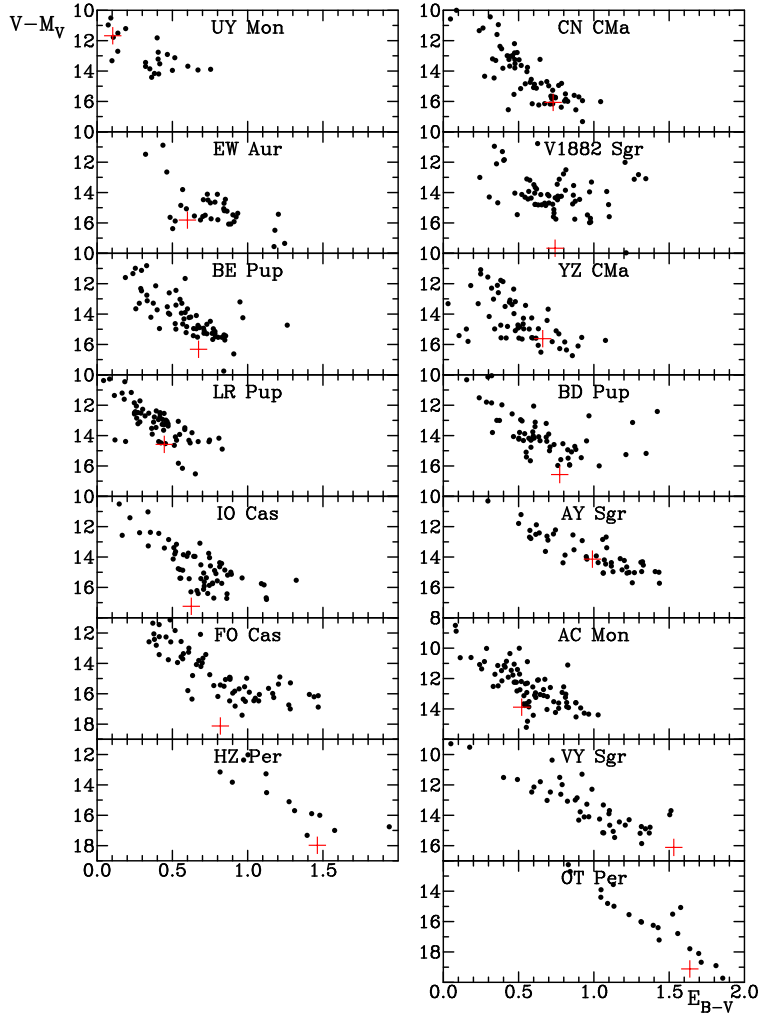}
\end{center}
\caption{\small{Variable-extinction diagrams for the fields examined in this study, identified by the Cepheid of interest. Filled symbols denote AF stars near each Cepheid, and red plus signs are ``predictions'' for the Cepheid parameters.}}
\label{fig3}
\end{figure*}

The additional scatter in Fig.~\ref{fig2} illustrates some of the problems associated with dereddening stars in {\it BV(RI)}$_C$ color-color diagrams. Such scatter for a small proportion of stars is a common characteristic of color-color diagrams, including those in {\it UBV} and those used for 2MASS photometry, and is readily explained in most cases by observational error, typically one or more of the magnitudes in the observed colors falling too close to the survey limits. At bright magnitude limits the typical stars encountered in Galactic star fields are A dwarfs and GK giants \citep{mc65}, but the demographic changes as the magnitude limits increase, and faint, nearby degenerate stars and M dwarfs begin to compound the picture with their unusual colors and larger measuring uncertainties. Emission, binarity, overly bright stars suffering from image saturation, objects lying away from the main sequence \citep{ca93}, and low quality observations for faint stars all combine to generate scatter in such plots, but the use of two separate color-color diagrams and limiting the analysis to stars fainter than the bright survey limits and brighter than the faint survey limits assures that likely AF dwarfs are identified properly. They are the small fraction of objects that deredden uniquely to the intrinsic relation for such stars in Fig.~\ref{fig2}, as confirmed by their {\it JHK}$_{\rm s}$ colors.

The derived reddenings for stars analyzed in each field were adjusted to equivalent color excesses for a B0 star according to their inferred intrinsic colors and the dependence of reddening on intrinsic color summarized by \citet{fe63}. Interstellar extinction affects the effective wavelengths of broad band {\it BV} filters differently according to the continuum of the star being observed, such that a typical Cepheid of, say, $(B-V)_0 = 0.60$ reddened by $E(B-V) = 0.90$ suffers identical extinction to a B0 dwarf star of $(B-V)_0 = -0.30$ reddened by $E(B-V) = 0.97$ \citep[cf. $\eta$ factor of][]{fe63}. Such adjustments are necessary when comparing reddenings of stars across a wide range of intrinsic color. Zero-age main sequence (ZAMS) values of $M_V$ as a function of intrinsic $(B-V)_0$ color \citep{tu76a,tu79} were also assigned to each star from the photometric dereddening solutions in order to provide estimates, or at least underestimates, of apparent distance modulus, {\it V--M}$_V$, for the stars in each field. The stars were then plotted in a variable-extinction diagram, such as those for each field summarized in Fig.~\ref{fig3}, as a means of assessing the likely space reddening of each Cepheid. Fig.~\ref{fig3} represents only a portion of each variable-extinction diagram, namely the most heavily-populated regions associated with the expected parameters for each Cepheid.

Mean $\langle B \rangle$ and $\langle V \rangle$ magnitudes are available for each Cepheid from \citet{bd07} and the present study, and rough estimates of reddening and luminosity were made from older, published period-color \citep[e.g.,][]{fe90a,fe90b} and period-luminosity \citep[e.g.,][]{tu92} relations, which do not differ substantially from more recent results \citep{tu01,tu10}. It was then possible to establish roughly where in each variable-extinction diagram the parameters for the Cepheid should fall, keeping in mind that an extremely liberal interpretation of such ``predictions'' is essential to avoid biasing the results. Some of the sample Cepheids are suspected Type II objects or overtone pulsators, for example, which affects estimates of both reddening and luminosity. Generous uncertainties of $\pm0.1$ or more in $E_{B-V}$ and $\pm 1$ in $M_V$ were therefore assumed in the analysis. Note that the ``predictions'' for IO Cas, FO Cas, and V1882 Sgr are inconsistent with expectations for classical Cepheids (they have larger predicted distance moduli than those for surrounding stars of similar reddening), which means that they are potential Type II objects. The situation for HZ Per, BD Pup, and OT Per is more ambiguous.

All such information is needed in order to establish what stars in the field of each Cepheid are useful for deriving its space reddening. In the case of UY Mon, for example, the estimated distance and reddening for the Cepheid associate it with the group of slightly reddened stars at $E_{B-V} \simeq 0.1$, ({\it V--M}$_V)_0 \simeq$ 11--12 ($d \simeq$ 1.6--2.5 kpc), but less distant and less reddened than stars of $E_{B-V} \simeq 0.5$, ({\it V--M}$_V)_0 \simeq$ 12.5--13 ($d \simeq$ 3--4 kpc), in its vicinity. The best match is therefore to the 4 stars within $5^\prime$ of UY Mon that are reddened by $E_{B-V} \simeq 0.1$. Other cases are more complex because of the patchy reddening that permeates most fields, but have been resolved by considering only stars close to each Cepheid.

\begin{figure}[!t]
\begin{center}
\includegraphics[width=7.0cm]{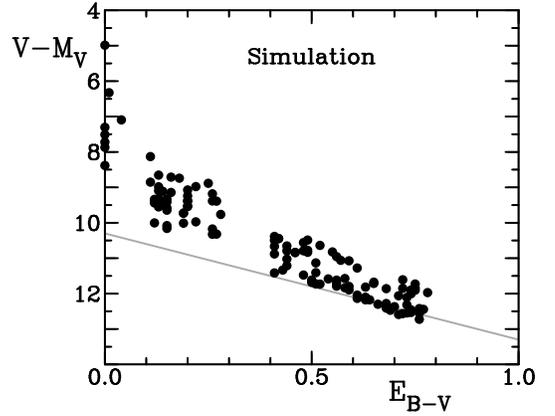}
\end{center}
\caption{\small{A simulated variable-extinction diagram for a star field where the reddening increases with distance according to dust clouds producing $E_{B-V}$ = 0.20, 0.50, and 0.70 located at distances of 0.5 kpc, 0.9 kpc, and 1.1 kpc, respectively, with associated uncertainties in $E_{B-V}$ of $\pm0.05$ and in $M_V$ of $\pm0.5$. The gray line denotes a standard reddening law of $R=A_V/E_{B-V}=3.0$ for stars at a distance of 1.1 kpc.}}
\label{fig4}
\end{figure}

What appears to be a continuous run of reddening with distance in some fields results from the patchy extinction in each field combined with larger-than-average uncertainties in the inferred color excesses, $E_{B-V}$. The extinction associated with most Galactic star fields is typically associated with individual dust clouds dispersed along the line of sight \citep{tu94}. But large uncertainties in reddening can confuse the picture. Fig.~\ref{fig4} presents the results of a simulation of such circumstances in a field where the scatter in the color excesses is taken to be $\pm0.05$, with associated uncertainties in absolute magnitude of $\pm0.5$, both quantities being applied randomly to the test points. Star densities were assumed constant as a function of distance, and the input parameters included specific amounts of reddening and extinction arising in discrete dust clouds located along the line of sight at distances of 0.5 pc, 0.9 pc, and 1.1 pc, producing mean color excesses of $E_{B-V}$ = 0.2, 0.5, and 0.7, respectively. An extinction law with $R=3.0$ is depicted for the last group. The parameters, in particular the adopted scatter, were chosen in order to produce the greatest complexity in the resulting variable-extinction diagram. The resulting scatter produces results similar to some of the variable-extinction diagrams of Fig.~\ref{fig3}.

\begin{figure*}[!t]
\begin{center}
\includegraphics[width=10.0cm]{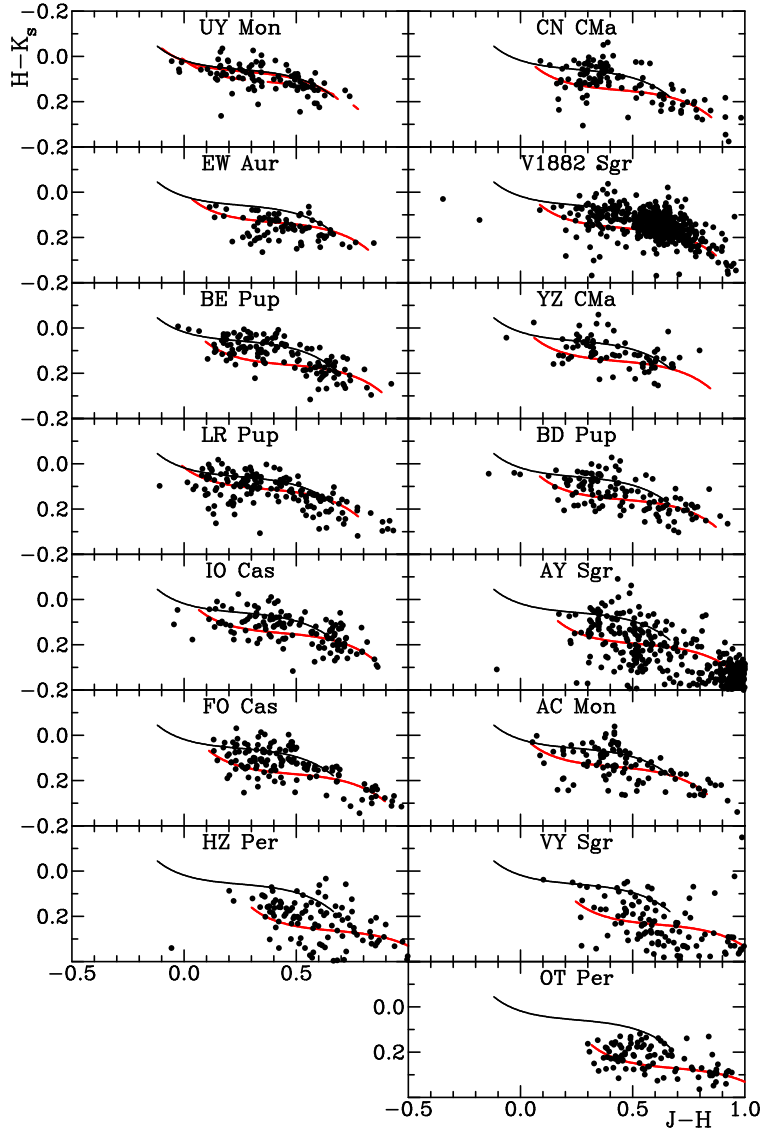}
\end{center}
\caption{\small{2MASS color-color diagrams, {\it H--K}$_{\rm s}$ versus {\it J--H}, for stars within $5^\prime$ of each Cepheid, from observations by Cutri et al. (2003). The intrinsic relation for main sequence stars is plotted as a black line, while red lines depict the adopted reddening for stars associated with the Cepheids, from column (7) of Table~\ref{tab2}. The dashed red line for UY Mon represents the alternate solution for that field.}}
\label{fig5}
\end{figure*}

The large range of inferred reddenings for field stars near each Cepheid was reduced further through an analysis of 2MASS observations \citep{cu03} for likely BAF-type stars in the same fields, analyzed in similar fashion to that employed by \citet*{te08} and \citet{tu11}. 2MASS observations were used separately without combination with the {\it BV(RI)}$_C$ photometry in order to avoid potential zero-point problems. The observed {\it J--H} and {\it H--K}$_{\rm s}$ colors for stars lying within $5^\prime$ of each Cepheid are shown in Fig.~\ref{fig5}, and were compared with the intrinsic relation for main-sequence stars in the 2MASS system \citep{tu11}, adjusted with a reddening slope ${\rm E}_{H-K}/{\rm E}_{J-H} = 0.49$ from \citet{tu11}. Multiple solutions in some cases, e.g., UY Mon and HZ Per, were resolved with reference to the variable-extinction diagrams and ``predicted'' values, and best fits were made by trial and error by establishing reddenings for which likely BAF-type stars had colors distributed randomly about the reddened intrinsic relation.

\setcounter{table}{1}
\begin{table*}[!t]
\caption[]{Space Reddenings for Cepheid Fields.}
\label{tab2}
\centering
\scriptsize
\begin{tabular*}{0.89\textwidth}{@{\extracolsep{-0.9mm}}lccccccccc}
\hline \noalign{\medskip}
Cepheid &$\log P$ &$E_{B-V}$(B0) &$E_{B-V}$(B0) &$E_{B-V}$(B0) &$E_{B-V}$(B0) &$E_{B-V}$ &$E_{B-V}$(B0) &No. &$E_{B-V}$(C$\delta$) \\
& &$<2^\prime$ &$<3^\prime$ &$<4^\prime$ &$<5^\prime$ &2MASS &Adopted &Stars &  \\
\noalign{\smallskip} \hline \noalign{\medskip}
\noalign{\smallskip}
UY Mon &0.380 &\nodata &$0.09$ &$0.09$ &$0.12 \pm0.01$ &$0.07 \pm0.07$ &$0.12 \pm0.01$ &4 &$0.11 \pm0.01$ \\
UY Mon$^1$ &\nodata &\nodata &$0.42$ &$0.44 \pm0.03$ &$0.43 \pm0.02$ &$0.37 \pm0.07$ &$0.43 \pm0.02$ &3 &\nodata \\
CN CMa &0.390 &$0.62 \pm0.02$ &$0.65 \pm0.02$ &$0.66 \pm0.02$ &$0.67 \pm0.02$ &$0.63 \pm0.12$ &$0.67 \pm0.02$ &11 &$0.63 \pm0.02$ \\
EW Aur &0.425 &$0.52$ &$0.63 \pm0.06$ &$0.61 \pm0.04$ &$0.61 \pm0.04$ &$0.53 \pm0.08$ &$0.61 \pm0.04$ &6 &$0.58 \pm0.03$ \\
V1882 Sgr &0.433 &$0.60 \pm0.01$ &$0.68 \pm0.03$ &$0.70 \pm0.02$ &$0.68 \pm0.02$ &$0.69 \pm0.05$ &$0.68 \pm0.02$ &20 &$0.64 \pm0.01$ \\
BE Pup &0.458 &$0.80$ &$0.72 \pm0.04$ &$0.69 \pm0.03$ &$0.68 \pm0.02$ &$0.73 \pm0.12$ &$0.68 \pm0.02$ &20 &$0.64 \pm0.02$ \\
YZ CMa &0.499 &\nodata &$0.71 \pm0.10$ &$0.64 \pm0.09$ &$0.60 \pm0.03$ &$0.61 \pm0.07$ &$0.60 \pm0.03$ &10 &$0.56 \pm0.03$ \\
LR Pup &0.523 &$0.44$ &$0.47 \pm0.02$ &$0.46 \pm0.01$ &$0.45 \pm0.01$ &$0.37 \pm0.07$ &$0.45 \pm0.01$ &14 &$0.42 \pm0.01$ \\
BD Pup &0.593 &$0.63 \pm0.03$ &$0.71 \pm0.03$ &$0.74 \pm0.03$ &$0.72 \pm0.02$ &$0.69 \pm0.08$ &$0.72 \pm0.02$ &19 &$0.67 \pm0.02$ \\
IO Cas &0.748 &$0.74$ &$0.64 \pm0.03$ &$0.62 \pm0.03$ &$0.63 \pm0.03$ &$0.63 \pm0.08$ &$0.63 \pm0.03$ &14 &$0.59 \pm0.02$ \\
AY Sgr &0.818 &\nodata &$1.04 \pm0.02$ &$1.02 \pm0.04$ &$1.00 \pm0.03$ &$0.97 \pm0.05$ &$1.00 \pm0.03$ &8 &$0.94 \pm0.02$ \\
FO Cas &0.832 &$0.99$ &$0.99$ &$0.83 \pm0.07$ &$0.82 \pm0.06$ &$0.78 \pm0.14$ &$0.82 \pm0.06$ &7 &$0.76 \pm0.05$ \\
AC Mon &0.904 &\nodata &$0.61$ &$0.61 \pm0.02$ &$0.59 \pm0.02$ &$0.56 \pm0.05$ &$0.59 \pm0.02$ &5 &$0.55 \pm0.01$ \\
HZ Per &1.052 &$1.39$ &$1.39$ &$1.48 \pm0.07$ &$1.48 \pm0.05$ &$1.42 \pm0.10$ &$1.48 \pm0.05$ &4 &$1.36 \pm0.04$ \\
VY Sgr &1.132 &\nodata &$1.50$ &$1.41 \pm0.04$ &$1.35 \pm0.04$ &$1.24 \pm0.15$ &$1.35 \pm0.04$ &3 &$1.24 \pm0.04$ \\
OT Per &1.416 &\nodata &$1.70$ &$1.52 \pm0.11$ &$1.52 \pm0.09$ &$1.47 \pm0.12$ &$1.52 \pm0.09$ &5 &$1.39 \pm0.08$ \\
\noalign{\smallskip} \hline
\end{tabular*}
\\{\smallskip}
$^1$ Alternate solution for stars of large reddening near UY Mon.
\end{table*}

The {\it JHK}$_{\rm s}$ colors generate independent $E_{B-V}$ reddenings for the stars in the Cepheid fields, with the advantage of being more closely tied to B-type stars in the fields, stars that lie blueward of the ``kink'' in the intrinsic relation and that may share a common origin with the Cepheid. The reddenings have slightly larger uncertainties because of larger photometric scatter in the observations (by factors of 3--5 relative to the {\it BV(RI)}$_C$ photometry) in combination with the correction from infrared to optical reddening. They are nevertheless crucial for narrowing the range of potential color excesses for some of the sample Cepheids, such as IO Cas. A few objects in the sample appear to be Type II Cepheids, but that does not affect the results. In most cases the implied reddening near the Cepheid is fairly evident.

\section{{\rm \footnotesize RESULTS}}
The results of the variable-extinction studies of the sample Cepheids are presented in Table~\ref{tab2}, which lists the inferred space reddening for each Cepheid derived from likely AF-type stars lying within different angular radii, from $2^\prime$ to $5^\prime$ distant, and the reddening inferred from 2MASS {\it JHK}$_{\rm s}$ colors for stars within $5^\prime$ of the Cepheid. The adopted color excess in each case, column (8), was the average for all AF-type stars lying within $5^\prime$ of the Cepheid and of comparable inferred distance, and the resulting reddening, which is equivalent to that for a star of spectral type B0, was converted in the last column of the table to a value appropriate for the derived intrinsic color of the Cepheid, again using the relationship of \citet{fe63}. The number of stars used to obtain the space reddening is listed in the second last column of Table~\ref{tab2}. In rich fields that number is of order $\sim$10--20, whereas in more poorly populated regions or fields with a larger spread in reddening, less than half a dozen reference stars were available. Two solutions are presented for UY Mon, for reasons indicated later, although the adopted solution for the Cepheid is the first.

\begin{figure}[!t]
\begin{center}
\includegraphics[width=7.0cm]{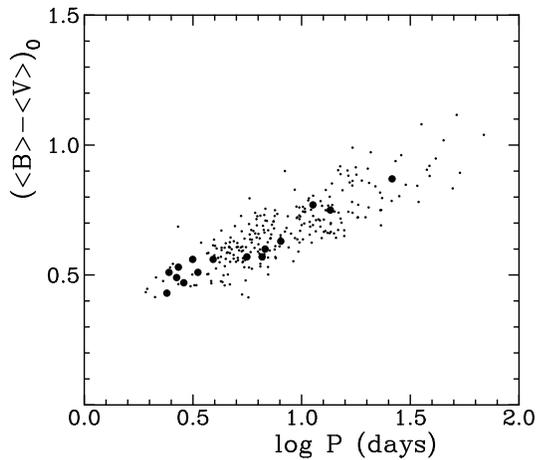}
\end{center}
\caption{\small{Derived intrinsic $\langle B \rangle$--$\langle V \rangle$ colors for sample Cepheids (large symbols) relative to other Cepheids with derived reddenings (small symbols, Turner 2001), as a function of pulsation period.}}
\label{fig6}
\end{figure}

It is worth noting that the 2MASS reddenings fall closer to the converted color excess for the Cepheid than to the B0-star reddenings. A reasonable explanation lies in the fact that the BAF-star reddenings in a 2MASS color-color diagram tend to be dominated by the AF stars, particularly likely F-type stars, since B-type stars are less common in the small fields analyzed and are potentially affected by circumstellar affects \citep{tu11}. The 2MASS reddenings are also important for providing confirmation of the relationship used to convert B0-star color excesses to those appropriate for a star with the average unreddened colors of the Cepheid, which are very similar to F-type dwarfs in the 2MASS samples.

The derived intrinsic colors are presented in Table~3, which also provides in column (3) the variable star type designation for the object from the {\it General Catalogue of Variable Stars} \citep{sa04}. Type II Cepheids are designated as CWB for short period (1--8 days) BL Herculis variables, and CWA for longer period (8--20 days) W Virginis stars. The intrinsic $(\langle B \rangle$--$\langle V \rangle)_0$ colors are plotted in Fig.~6 relative to a set of similar intrinsic colors calibrated relative to other Cepheids with field reddenings (binary, cluster or association members) and Cepheids with reddenings derived using reddening-free indices \citep{tu01}.

All of the program stars studied here appear to have intrinsic colors consistent with those of classical Galactic Cepheids, despite remaining questions about the true population status for some of them. The spread in color at a given value of pulsation period {\it P} for the reference sample can be attributed to the natural width of the Cepheid instability strip, and in a few cases to erroneous reddenings. The location of individual Cepheids in our sample relative to the hot (blue) and cool (red) edges of the strip, if used in conjunction with light amplitude and rate of period change, provides useful information about what stage the Cepheids have reached in their evolution \citep*{te06}.

\begin{figure}[!t]
\begin{center}
\includegraphics[width=7.0cm]{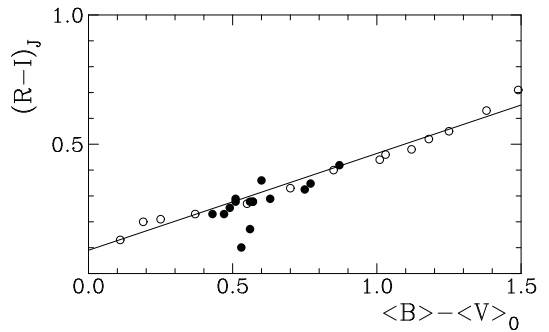}
\end{center}
\caption{\small{Derived Johnson system ({\it R--I})$_J$ colors for sample Cepheids (filled circles) relative to intrinsic $(\langle B \rangle$--$\langle V \rangle)_0$ colors, with derived colors for AFGK supergiants from Johnson (1966) plotted for reference purposes as open circles. The line is a fitted relation for the supergiants. The anomalous Cepheids are V1882 Sgr and YZ CMa.}}
\label{fig7}
\end{figure}

\setcounter{table}{2}
\begin{table}[!t]
\caption[]{Unreddened {\it BV(RI)}$_C$ Colors for Sample Cepheids.}
\label{tab3}
\centering
\scriptsize
\begin{tabular*}{0.46\textwidth}{@{\extracolsep{-2.0mm}}lcccccc}
\hline \noalign{\medskip}
Cepheid &$\log P$ &Type &$\langle B \rangle$--$\langle V \rangle$ &{\it V--R}$_C$ &{\it V--I}$_C$ &Deduced  \\
& & & & & &Type \\
\noalign{\smallskip} \hline \noalign{\medskip}
\noalign{\smallskip}
UY Mon &0.380 &DCEPS &+0.43 &+0.28 &+0.51 &DCEPS$^1$ \\
CN CMa &0.390 &CWB: &+0.51 &+0.33 &+0.61 &DCEP \\
EW Aur &0.425 &DCEP &+0.49 &+0.29 &+0.54 &DCEP \\
V1882 Sgr &0.433 &CEP &+0.53 &+0.30 &+0.42 &CWB \\
BE Pup &0.458 &CWB: &+0.47 &+0.29 &+0.52 &DCEP \\
YZ CMa &0.499 &CWB: &+0.56 &+0.28 &+0.46 &CWB? \\
LR Pup &0.523 &CEP &+0.51 &+0.32 &+0.59 &DCEP \\
BD Pup &0.593 &DCEP &+0.56 &+0.36 &+0.63 &DCEP \\
IO Cas &0.748 &DCEP &+0.57 &+0.36 &+0.63 &CWB? \\
AY Sgr &0.818 &DCEP &+0.57 &+0.31 &+0.58 &DCEP \\
FO Cas &0.832 &DCEP &+0.60 &+0.44 &+0.78 &CWB? \\
AC Mon &0.904 &DCEP &+0.63 &+0.35 &+0.63 &DCEP \\
HZ Per &1.052 &DCEP &+0.77 &+0.40 &+0.73 &DCEP \\
VY Sgr &1.132 &DCEP &+0.75 &+0.43 &+0.74 &DCEP \\
OT Per &1.416 &DCEP &+0.87 &+0.49 &+0.88 &DCEP \\
\noalign{\smallskip} \hline
\end{tabular*}
\\{\smallskip}
$^1$ Fundamental mode pulsation likely.
\end{table}

Fig.~\ref{fig7} plots the inferred ({\it R--I})$_J$ colors for our sample Cepheids relative to their $(\langle B \rangle$--$\langle V \rangle)_0$ colors, transposed from ({\it R--I})$_C$ using the relationships of \citet{fe83}. Included are inferred colors for AFGK supergiants from \citet{jo66}, in similar fashion to the analysis of \citet*{de78}. There is good agreement of the derived intrinsic colors for sample Cepheids with the colors expected for supergiant stars, with the exception of V1882 Sgr and YZ CMa. Both objects have faint optical companions that may contaminate the {\it (RI)}$_C$ photometry, so the results of Fig.~\ref{fig7} do not necessarily indicate a significant difference in color between classical and Type II Cepheids. The data also indicate the general success of the procedure adopted in this study. The temperature spread of the instability strip is less marked in intrinsic ({\it R--I})$_J$ and ({\it R--I})$_C$ colors than in $(\langle B \rangle$--$\langle V \rangle)_0$ color, as expected.

With regard to the variable star designations for specific objects in Table~\ref{tab3}, the short period Cepheids UY Mon and CN CMa are solar-metallicity stars according to \citet{di90}, who derived reddenings of $E_{B-V} = 0.15\pm0.08$ and $0.61\pm0.10$, respectively, for the stars from Walraven photometry. Both values are consistent with the present results, although of lower precision. UY Mon is an s-Cepheid, its light curve being indistinguishable from a sine wave, many of which are overtone pulsators, yet fundamental mode pulsation provides the simplest solution to the variable-extinction study presented here (Fig.~3), provided it is a classical Cepheid. CN CMa is a suspected Type II Cepheid because of its unusually large amplitude for a short-period pulsator, but that designation is not consistent with its photometric solar metallicity \citep{di90} or the variable-extinction analysis. Type II Cepheids are known to display a wide range of abundances, from solar to below solar \citep{ha81}, so that is not necessarily a good criterion to decide Population type for CN CMa. A better case for it being a classical Cepheid lies in the variable-extinction results (Fig.~\ref{fig3}). Its {\it (RI)}$_C$ and {\it (RI)}$_J$ colors are also consistent with those of a classical Cepheid (Fig.~\ref{fig7}), which is the designation preferred here.

For four of the sample objects the field star variable-extinction data produce results indicating that the Cepheid's luminosity may have been overestimated in the analysis, i.e., it might be a Type II object. The Cepheids are V1882 Sgr, BE Pup, IO Cas, and FO Cas. The latter two objects are identified as classical Cepheids in the GCVS, so either the variable-extinction results for them are anomalous or the classification of the stars is erroneous. The latter possibility is suspected here, so their deduced types as Type II Cepheids in the last column of Table~\ref{tab3} reflect the results of the variable-extinction analysis. V1882 Sgr has the characteristic light curve of a Cepheid, but is of uncertain type. The variable-extinction results suggest that it is a Type II object, while its {\it V(RI)}$_C$ colors (Table~\ref{tab3}) are slightly bluer than those of classical Cepheids in the sample, which may indicate a metal-poor object. The ``CEP" designation for the star in the GCVS is presumably preliminary, and it appears likely to be a Type II object. Similarly, the variable-extinction results for BE Pup are consistent with either a classical or Type II Cepheid, while its {\it V(RI)}$_C$ colors (Table~\ref{tab3}) are similar to those of classical Cepheids in the sample. BE Pup is also a double-mode pulsator \citep{wo04}, which is suggestive of a classical Cepheid, our preferred designation.

YZ CMa is suspected to be a classical Cepheid according to the variable-extinction analysis, but its {\it V(RI)}$_C$ colors are also slightly bluer than those of other classical Cepheids. The ``CWB:" designation in the GCVS may therefore be correct, although further study is needed to resolve the question. In general, the studies summarized here consistently tend to identify most sample objects (11/15) as classical Cepheids. Two of the likely Type II objects, V1882 Sgr and YZ CMa, appear slightly bluer in {\it V--R}$_C$ and {\it V--I}$_C$ than other Cepheids in the sample, but the photometry for both stars may suffer from contamination by close companions, so their distinctive difference in color cannot be considered definitive. Detailed atmospheric abundance studies of both stars should help to solidify their classifications and establish if there are obvious differences in color between classical and Type II Cepheids.

\begin{figure}[!t]
\begin{center}
\includegraphics[width=7.0cm]{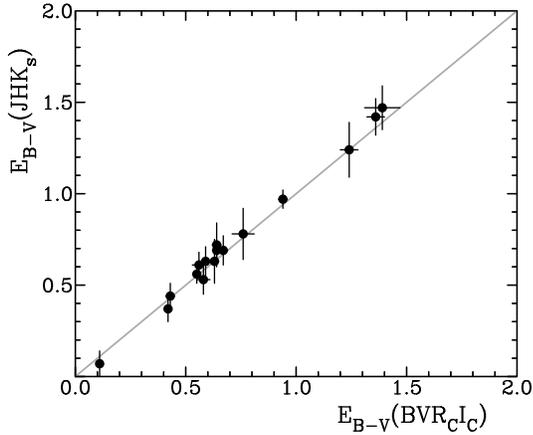}
\end{center}
\caption{\small{Comparison of derived 2MASS reddenings, with their uncertainties, for the Cepheids in our sample with the color-corrected broad band {\it BV(RI)}$_C$ reddenings. The gray line represents expectations for an exact match.}}
\label{fig8}
\end{figure}

Fig.~\ref{fig8} plots the inferred 2MASS reddenings for our sample Cepheids relative to their broad band {\it BV(RI)}$_C$ reddenings, where we have included the second result for the UY Mon field because of the obvious separation in Figs.~\ref{fig3}~and~\ref{fig5} of stars of small reddening from those of large reddening, for which separate solutions were obtained. The close agreement of both sets of reddenings in Fig.~\ref{fig8} indicates that the two techniques are of comparable reliability for establishing field reddenings of Galactic objects, although there is clearly greater precision in the {\it BV(RI)}$_C$ color excesses.

\section{{\rm \footnotesize DISCUSSION}}
The present study was undertaken as a feasibility test of the use of CCD {\it BV(RI)}$_C$ photometry for studying the space reddening of Galactic Cepheids. The {\it BV(RI)}$_C$ data by themselves were sometimes insufficient to define the field reddenings of each Cepheid unambiguously, but the addition of 2MASS reddenings and initial ``predictions'' guided the process successfully. The number of Galactic Cepheids of well-established field reddening has been increased from 40 \citep{lc07} to 55 by this study, and the sample now includes a few Type II Cepheids, some of which appear to display slightly bluer intrinsic {\it (V--R)}$_C$ and {\it (V--I)}$_C$ colors relative to their classical Cepheid cousins. A comparable test using reddenings derived from 2MASS {\it JHK}$_s$ colors indicates that the 2MASS survey may also be suitable for the derivation of space reddenings, for a variety of stellar types and not just Cepheids. Further studies of that possibility have already been initiated \citep*{me08a,me08b}.

The results of this study have already been used in a preliminary mapping of the Cepheid instability strip using Cepheid reddenings tied to space reddenings and spectroscopic reddenings of Galactic Cepheids \citep{tu01}, the results of which can be seen in the data plotted in Fig.~\ref{fig6}. The addition of {\it U}-band observations would have assisted the analysis considerably, but it is important to note that accurate Johnson system {\it U}-band observations are difficult to obtain with some CCD/filter combinations \citep[see comments by][]{tu11}. Despite that, the success of the present study using both {\it BV(RI)}$_C$ and {\it JHK}$_s$ photometry should spur further investigations that take advantage of existing survey photometry to study the reddening in interesting Galactic star fields.

\subsection*{{\rm \scriptsize ACKNOWLEDGEMENTS}}
\small{The present study was supported in part by research funding awarded through the Natural Sciences and Engineering Research Council of Canada (NSERC), through the Russian Foundation for Basic Research (RFBR), and through the program of Support for Leading Scientific Schools of Russia. This publication makes use of data products from the Two Micron All Sky Survey, which is a joint project of the University of Massachusetts and the Infrared Processing and Analysis Center/California Institute of Technology, funded by the National Aeronautics and Space Administration and the National Science Foundation.}

\end{document}